\documentclass[12pt]{article}   
\usepackage{epsfig,times}  

\begin{document}
\thispagestyle{empty} 


 \renewcommand{\topfraction}{.99}      
 \renewcommand{\bottomfraction}{.99} 
 \renewcommand{\textfraction}{.0}


\newcommand{\nc}{\newcommand}

\nc{\qI}[1]{\section{{#1}}}
\nc{\qA}[1]{\subsection{{#1}}}
\nc{\qun}[1]{\subsubsection{{#1}}}
\nc{\qa}[1]{\paragraph{{#1}}}

\def\qbu{\hfill \par \hskip 6mm $ \bullet $ \hskip 2mm}
\def\qee#1{\hfill \par \hskip 6mm #1 \hskip 2 mm}

\nc{\qfoot}[1]{\footnote{{#1}}}
\def\qL{\hfill \break}
\def\qpar{\vskip 2mm plus 0.2mm minus 0.2mm}
\def\qtvi{\vrule height 2pt depth 5pt width 0pt}
\def\qth{\vrule height 12pt depth 0pt width 0pt}
\def\qtb{\vrule height 0pt depth 5pt width 0pt}
\def\tvi{\vrule height 12pt depth 5pt width 0pt}

\def\qparr{ \vskip 1.0mm plus 0.2mm minus 0.2mm \hangindent=10mm
\hangafter=1}

\def\qdec#1{\par {\leftskip=2cm {#1} \par}}

\def\qdpt{\partial_t}
\def\qdpx{\partial_x}
\def\qddpt{\partial^{2}_{t^2}}
\def\qddpx{\partial^{2}_{x^2}}
\def\qn#1{\eqno \hbox{(#1)}}
\def\qds{\displaystyle}
\def\qw{\widetilde}
\def\qmax{\mathop{\rm Max}}   
\def\qmin{\mathop{\rm Min}}   


\def\qci#1{\parindent=0mm \par \small \parshape=1 1cm 15cm  #1 \par
               \normalsize}

\null
\vskip 1.5 cm

\centerline{\bf \Large Suicide: the key role of short range ties}

\vskip 1cm
\centerline{\bf Bertrand M. Roehner $ ^1 $ }
\vskip 4mm
         
\centerline{\bf Institute for Theoretical and High Energy Physics}
\centerline{\bf University Paris 7 }

\vskip 2cm

{\bf Abstract}\quad 
The paper explores the connection between short-range
social ties (i.e. links with close relatives) and the
occurrence of suicide. The objective is to discriminate between
a model based on social ties and a model based on
psychological traumas.
Our methodological strategy is to
focus on instances characterized
by the severance of some social ties. We consider
several situations of this kind.
(i) Prisoners in the first days after their incarceration.
(ii) Prisoners in solitary confinement
(iii) Prisoners who are transferred from one prison to another.
(iv) Prisoners in closed versus open prisons.
(v) Prisoners in the weeks following their release.
(vi) Immigrants in the years following their relocation. 
(vii) Unmarried versus married people. 
\qpar

Furthermore,
in order to test the impact of major shocks we consider the
responses in terms of suicides to the following shocks.
(i) The attack of September 11, 2001 in Manhattan.
(ii) The Korean War.
(iii) The two world wars.
(iv) The Great Depression in the United States.
(v) The hyperinflation episode of 1923 in Germany.
Major global traumatic shocks
such as 9/11 or wars have no influence on suicide rates once
changing environment conditions
have been controlled for.
\qpar

Overall, it turns out that the observations 
have a natural interpretation in terms of
short-range ties. 
In contrast, the trauma model seems unable to adequately
account for many observations.
\vskip 1cm

\centerline{January 31, 2005}

\vskip 8mm
\centerline{\it Preliminary version, comments are welcome}

\vskip 1cm
Key-words: social ties, family ties, suicide, prison, wars. 
\vskip 1cm

1: Bertrand Roehner, LPTHE, University Paris 7, 2 place Jussieu, 
F-75005 Paris, France.
\qL
\phantom{1: }E-mail: roehner@lpthe.jussieu.fr
\qL
\phantom{1: }FAX: 33 1 44 27 79 90

\vfill \eject

\qI{Introduction}

The paper examines two alternative interpretations of the
phenomenon of suicide. The first one was put 
forward in a previous paper (Roehner 2005) while the second one
was embodied in objections raised by this paper.
The two interpretations are recalled in the introduction.
The rest of the paper presents evidence
specifically selected in order to provide ``critical experiments''
which should allow a clear choice to be made
between the interpretations in competition.

In Roehner (2005), a paper which for the sake of concision will be referred
to as {\it Bridge}, a parallel was drawn between the strength of intermolecular
bonds on one hand and of interpersonal ties on the other hand. It was
pointed out that, irrespective of the precise nature of the bonds, whether
ion-ion, dipole-dipole, dipole-induced dipole, etc., it is their {\it strength} 
which explains a broad range of phenomena such as for instance boiling
point, heat of vaporization, rate of evaporation, etc. Similarly, it
was observed that, irrespective of their precise nature, whether
husband-wife, parent-children, friend-friend and so forth,
it is the strength of these ties which determine a broad range of
social phenomena such as for instance school dropout, desertion, suicide,
etc. 
In {\it Bridge} I suggested that, as we do not yet have adequate means for
measuring the strength of social ties, one should focus on situations
in which social ties have been severed. Imprisonment, emigration and
divorce are three cases where former ties are severed. In the case of
imprisonment all former links with family, friends or neighbors are cut off.
In the case of emigration, only the links with the family members who 
emigrated together are preserved. In the case of divorce, the bond
between husband and wife is broken; for one of the partner (usually the 
former husband) the links with the children are also weakened to a large extent.
\qpar

I now come to the alternative interpretation as suggested by
Professors 
Bernard Diu (oral communication) and Didier Sornette (email of May 16, 2004).
The point is that the suicides
occurring shortly after an arrest (which was one of the situations
on which {\it Bridge} focused) can also be attributed to the psychological
trauma of being arrested and incarcerated. Naturally, the same argument
can be used in the two other cases as well. Instead of the severance
of the social ties one would 
incriminate the shock and grief of the separation. For short, this 
perspective will be referred to as the trauma interpretation. 
Does it really matter
whether one adopts one interpretation or the other?
It does make a difference for at least two reasons.
(i) The social tie framework is not dissimilar to what can be observed
in physical or biological systems. On the contrary, the trauma mechanism
relies on psychological reactions for which there is no parallel in the 
physical or biological worlds. If we settle for this interpretation we must
definitely give up the prospect of a unified perspective. 
(ii) Even though we do not yet have any means for measuring 
the strength of social ties
one can expect that this will become possible in the future. Once
we will be able to measure the coupling strength between
(for instance) husband and wife, one may expect that the {\it same 
coupling} will play a role (albeit not exactly the same role) in 
all phenomena in which this link is dissolved as for instance in the
death, imprisonment or separation of one of the partners.
On the contrary, in the trauma framework one
has to resort to different (and largely {\it ad hoc})  shocks 
which have no reason to be
related in any way. In short, the trauma
perspective is hardly conducive to the building of a satisfactory 
theoretical framework.
\qpar

In order to prevent any misinterpretation we must insist on the importance 
of the
{\it transient state} between two equilibrium situations.
When the social
environment of an individual changes as in the case of imprisonment or
emigration, the old ties are severed and it takes some time to establish
new links even if the new environment is no less favorable than the old
one. Thus, for instance, if a prisoner is moved from one prison to
another, it will require some time to establish ties in the new 
establishment even if the latter is very similar to the previous one. 
This phenomenon is very similar to what happens when two liquids
$ A $ and $ B $ are mixed. There is a transient state during which new
$ A-B $ bonds are established between the molecules of the two liquids. 
Naturally, at the molecular level the characteristic time of bond
reorganization is very short ($ \sim 10^{-9}\  \hbox{second} $) but before
a new $ A-B\ $ bond can be established the molecules must be close 
enough and this may take longer (it mainly depends upon the steering device).
\qpar

The paper proceeds as follows. The second section provides evidence
about the sharp increase in suicide rates which occurs consecutively
to imprisonment, or consecutively to a change in imprisonment conditions.
In the third section, I compare suicide rates for
immigrants versus non-immigrants. In the fourth section I compare
suicide rates for
married versus non-married individuals. 
In the fifth section I show that macro-traumas such as wars
or the attack of September 11, 2001 have no impact on suicide rates. 
All these cases
have been selected with the purpose of discriminating between the
two alternative interpretations. 
\qpar

Before we begin, we should explain how this study
is related to the work centered on stock markets which was done
by econophysicists in the last decade. If momentarily we
forget the economic
agents who buy and sell stocks, we can then see them as entities which
move in a one-dimensional space defined by their prices. Their moves
obviously are interdependent in the sense that closely related sectors
(e.g. microprocessors, computers or softwares) move in parallel. This explains
why the 
question of the interdependence between stocks is of central importance.
Econophysics groups have been using various approaches to measure
couplings between stocks. 
One of these is to
analyze the properties of the correlation matrix of
a broad sample of stock prices%
\qfoot{This approach is at the base of Markowitz's portfolio theory.}%
,
but there are several others. The interested
reader can find more information in the following papers: Bonano et al. (2001),
Drozdz et al. (2001), Kim et al. (2004), Mantegna et al. (2004), Plerou et
al. (2001), Sornette et al. (2003), Stauffer et al. (1999). The question of
interaction strength is not only of importance in finance, but also in
economics. In this respect on can
mention the exploration of the coupling strengths between several
spatially separated commodity markets (see for instance Roehner 1995). 
In all  these
cases the links can be identified  by probing the interdependence of
prices. This raises an obvious question:
for social ties do we have something similar to prices?
If the thesis developed in this paper is correct, suicide rates can indeed
be used  for the purpose of measuring social ties just as prices permit to
estimate economic ties.

\qI{Suicide in prison}

The purpose of Fig. 1 is to summarize some of the evidence presented
in {\it Bridge}. These data describe the evolution in suicide rate in the
hours, days, weeks and years following an incarceration. The decrease
roughly follows a power law at least until the plateau phase is reached.

  \begin{figure}[htb]
   \centerline{\psfig{width=15cm,figure=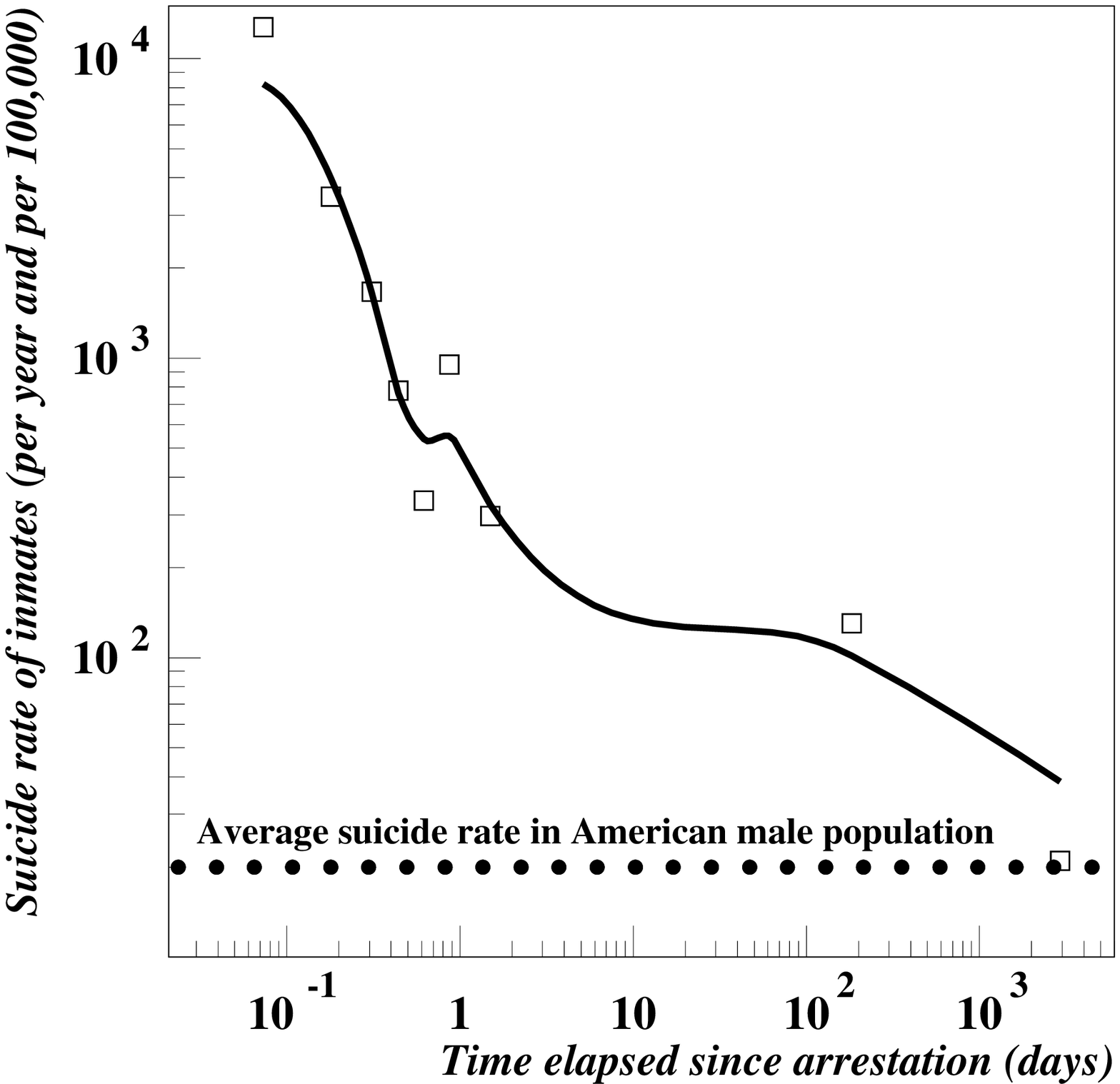}}
    {\bf Fig.1: Suicide rate of prisoners in the course of time after their
incarceration.}
{\small In the two days following incarceration the suicide rate decreases
rapidly (the relaxation time is 6.2 hours). Subsequently, the decrease
continues at a much slower rate (in the first week the relaxation time
is 81 hours). After two years the suicide rate tends towards the
rate in the male population}.
{\small \it Sources: Hayes et al. (1988, p. 36), Roehner (2005)}.
 \end{figure}

The exponent of the power law is $ 1.1 $, in other words:
$$ \hbox{suicide rate} \sim 1/
\hbox{[time elapsed since incarceration]}^{1.1} $$

Over the first week the average relaxation time is about 2 days. 
In itself this phenomenon does not allow us to discriminate between the
two interpretations%
\qfoot{However, it will be useful by providing a broad
overview of typical orders of magnitude to which subsequent suicide
rates can be compared.}%
.
In the tie interpretation one would say that the propensity
to commit suicide declines as new links are formed for instance with
fellow inmates, police officers, wardens or layers. 
In the trauma perspective
one would say that the trauma brought about by the incarceration
recedes in the course of time in the minds of the inmates. One possible
way for discriminating between the two explanations would be to identify
a phenomenon which results in a sharp increase of suicide rates without
involving an initial trauma. Another test would be to study a situation
which involves an initial trauma but does {\it not} result in increased suicide
rates. In what follows, these two tests will be used successively. 

\qA{Impact of solitary confinement}
The evidence on which we rely in this paragraph concerns suicides
which occurred in solitary confinement cells in New York State prisons.
These cells, also called ``special housing units'', are used for the purpose
of punishing inmates for rebellious behavior. The regime in these cells
implies confinement 23 hours a day and little human contacts with the outside.
On December 31, 2001 
there were $ 3,654 $ inmates in solitary confinement cells representing
$ 5.4\% $ of the total population of inmates in New York State. The average
time spent in these cells was $ 4.6 $ months in 2001. Under 
New York State law there
is no limit to the amount of time an inmate can be kept in confinement.
In 2001 the average annual suicide rate in prison was $ 26 $ per $ 100,000 $, but
it reached $ 140 $ (i.e. $ 5.4 $ times higher) in solitary confinement. It is true
that the number of suicides in a single year was small ($ \sim 6 $) but a
a ratio of $ 5.4 $ is nevertheless highly significant%
\qfoot{The argument goes as follows. The expression of the standard
deviation $ \sigma _{m} $ of the mathematical expectation $ m $ of the suicide
rate is: $ \sigma _{m} =\sqrt{D/n} $, where $ D $ is the variance of the
suicide rate and $ n $ the number of inmates (Ventsel 1973, p. 307).
To
get an estimate for $ \sigma _{m} $ one first needs an estimate for $ D $.
It can be obtained from a time series of suicide rates. By taking 
for instance the county of Chautauqua in New York State (which has
about 65,000 male inhabitants) over 20 years (1979-1998)
one gets $ D=36 $ and $ \sigma _{m} = 1.3 $ for the total population
of $ 65,000 $ inmates in New York State. 
For the inmates in confinement the
$ \sigma _{m} $ will be $ \sqrt{1/0.054}=4.2 $ times larger, which gives
$ \sigma _{m} = 5.5 $. Thus, the difference 
$ 140-26=114 $ of suicide rates in confinement versus in prison
represents $ 20 $ times the standard deviation. The probability
of occurrence of a statistical fluctuation of this magnitude is exceedingly
small. For $ x=20 $ one gets:
$ 1-\Phi ^*(x)=(1/\sqrt{2})\hbox{erfc}(x/2) \sim (\sqrt{2/\pi })\exp (-x^2/4/x
\sim 10^{-44} $).}%
. 
Furthermore,
the same observation also holds for the 3-year period 
1998-2000: in this case the suicide rate in confinement was $ 8.7 $ times higher
than in prison. Over the four years 1998-2001 the average suicide rate 
in confinement is about $ 8 $ times higher than in prison. Let us now examine
the possible explanations of this effect in the two perspectives. In the tie 
interpretation, the higher suicide rate would be attributed to the severance
of the links with former inmates and wardens
and the difficulty of building new ones
due to confinement. 
In the trauma
interpretation, one would argue that the shift from prison to confinement
brought about a trauma which accounts for the higher suicide rate.
According to this interpretation the suicide rate in confinement should 
decrease in the course of time as the trauma subsides. If one could
know the distribution over time of the suicides which occur 
in confinement cells, that would provide a crucial tests. Indeed, as
few new ties are formed in confinement the tie hypothesis
would lead us to predict that the suicide rate in confinement should
decrease at a much slower rate than what we observed in Fig. 1. 
So far, we were not able to find such data.
\qpar

\qA{The silent system} 
A related effect
was observed in the 19th century. Starting from the premise
that prisoners learn criminal ways from each other and that on the
contrary isolation would put prisoners face to face with their conscience
and encourage penitential reflection,
prison reformers introduced what was called the {\it silent system}.
In the Pennsylvania system, prisoners
were kept isolated in individual cells all the time.
In the Auburn (New York State)
system they
were kept in cells during night and evening; 
during the day they had to work together but in enforced silence. 
In 1842 
the English writer Charles Dickens was authorized to visit the Pennsylvania's
penitentiary. In his account of the visit he wrote: ``A prisoner sees the
prison officers, but with that exception he never looks upon a human
countenance or hears a human voice. He is a man buried alive''
(Dickens 1842).
Under such circumstances the tie interpretation would lead  us to
expect high suicide rates%
\qfoot{Naturally, such a prediction is conditioned by the availability
of means for committing suicide. If there are not sheets, towels, glasses,
windows nor anything sharp that could be used as a knife, committing suicide
becomes nearly impossible. Naturally, living in a cell deprived of
almost everything would be a nightmare that is likely to affect the prisoner's
mental health. Under such circumstances, he may 
commit suicide not in the confinement cell but as soon as allowed
to leave it.
It is the difficulty of controlling for all these factors which makes 
the task difficult.}%
. 
This was indeed the case although
there is only scant quantitative evidence. 
\qbu  In a lecture given by the prominent British lawyer
Louis Blom-Cooper (1987),  we learn
that in 1877, after adoption of the silent system in Britain, ``the suicide
rate among prisoners was as high as $ 1,760 $ per $ 100,000 $''. 
\qbu Going into the
same direction are the high levels of suicide rates in prison mentioned in 
{\it Bridge} (table 2).  For instance $ 190/10^5 $ in Belgium (1872) and 
$ 860/10^5 $ in Saxony (1872), two places where we know
the silent system was adopted.
\qbu At a more qualitative level, we read that the Pennsylvania system
``led to high rates of suicide and insanity'' (Telzrow 2002)
\qfoot{A dissenting note is introduced by Dickens. In the
account of his visit, he writes that ``suicide rates are rare among
these prisoners''. Clearly, additional data for suicide rates in U.S. prisons
in the 19th century would be welcome.}%
.
\qA{Effects of a reorganization of social ties}
As already noted a change in the social environment results in
a disruption followed by a reorganization of social ties. This
effect is fairly similar to a trauma effect with the result that the
two effects can be easily mistaken one for the other. It is therefore
important to get detailed information.
Prison and army statistics offer useful quasi-experiments.

{\bf The test of remand centers}\quad
A first test
is offered by remand centers in Britain. Prisoners who are waiting to
be sentenced but were remanded in custody by the courts stay in
the remand areas of local jails. Because these remand centers are
used for short stays their population is in majority composed
of ``new'' prisoners. The following table shows that the transient state
which characterizes these prisoners results in suicide rates which are
about three times as high as the average prison rate. 

$$ \matrix{
\tvi 
\hbox{Type of prison} \hfill & 
\hbox{Annualized suicide rate (per 100,000} \cr
\qtb
\hbox{} \hfill & \hbox{of average daily population), 1996-1998} \cr
\noalign{\hrule}
\qth 
\hbox{Remand centers (stays shorter than 6 months)} & 389\cr
\qtb 
\hbox{Total prison population} \hfill & 119 \cr
\noalign{\hrule}
} $$
Source: Marshall et al. (2000)
\qpar

This observation is interesting in two respects.
(i) The fact that the suicide rate in remand centers is higher than in
solitary confinement is at first surprising. Probably the simplest explanation
is that prisoners have much more means of committing suicide in
remand centers than in solitary confinement cells. 
(ii) The analogues in the U.S. of the British remand centers are
the jails. But whereas in remand centers stays are limited to
6 months, in jails they are limited to one year. This is probably the reason
why suicide rates in remand centers are twice as high as in jail (for 
suicide rates in jail 
see the data given in {\it Bridge}, table 2, cases 3 to 6).
\qpar

{\bf Moving from one prison to another}\quad  In a study done by
MacDonald and Sexton (2002) we learn that suicide rates are increased
{\it every time} a prisoner is moved from one prison to another. If the time
is  counted from the day a prisoner arrives in a prison, the data show
that there are approximately as many suicides during the first 6 days
as during the following 53 days. The observation should be considered
as fairly robust because it relies on a set of data which covers 
the 11 years 1990-2000. In other words the annualized suicide rate is
about 9 times higher in the first 6 days as in the two following months.
Unfortunately, this observation does not distinguish between initial
incarceration (for which there is a very high suicide rate as we have
seen in Fig.1) and subsequent relocations. 
\qpar

{\bf Release from prison}\quad  When a prisoner is released there is
a transient state as well. Table 1a shows that the standard mortality
rate of released prisoners in the first week after release is
about 4 times as high as in prison. At first sight this could seem
as a perfect argument against the trauma interpretation.
Indeed, normally one would not interpret the fact of being released
as a trauma, quite the contrary. In the tie perspective
the surge in the first week 
has a natural interpretation as being caused by a transient
readjustment of social ties. However, we will see below that it would
be inappropriate to jump too quickly to definitive conclusions.
\qpar


\begin{table}[htb]

 \small 

\centerline{\bf Table 1a\quad Standardized mortality rates in prison
and after release (1996-1997)}

\vskip 3mm
\hrule
\vskip 0.5mm
\hrule
\vskip 2mm

$$ \matrix{
\tvi 
 \hbox{Situation}  \hfill & \hbox{Suicide:}  \hfill & \hbox{Accident:} \hfill \cr
 \hbox{}  \hfill & \hbox{standardized mortality rate}  \hfill &
 \hbox{standardized mortality rate} \hfill \cr
 \hbox{}  \hfill & \hbox{with respect to}  \hfill &
 \hbox{with respect to}  \hfill \cr
 \qtb \hbox{}  \hfill & \hbox{general population [=100]}  \hfill &
 \hbox{general population [=100]}  \hfill \cr
\noalign{\hrule}
\qth 
 \hbox{Prisoners in prison}  \hfill & \phantom{4,}860 & \phantom{2,}150 \cr
\hbox{Released prisoners, first week after release}  \hfill & 4,500 &  2,100 \cr
\qtb 
\hbox{Released prisoners, 6-month period after release}  \hfill &  1,140 &  
\phantom{2,}700\cr
\noalign{\hrule}
} $$

\vskip 1.5mm
Notes: The tables 1a,b are based on a 
sample of 233 deaths of ex-prisoners under community supervision.
\qL
Source: Sattar (2001, p. 47-49)
\vskip 2mm

\hrule
\vskip 0.5mm
\hrule

\vskip 10mm
\centerline{\bf Table 1b\quad Number of deaths per week of ex-prisoners
in the weeks after their release.}

\vskip 3mm
\hrule
\vskip 0.5mm
\hrule
\vskip 2mm

$$ \matrix{
\tvi 
\hbox{Situation}  \hfill & \hbox{Suicide:}  \hfill & \hbox{Natural causes:} \hfill &
\hbox{Accidents:} \hfill \cr
\hbox{}  \hfill & \hbox{number of deaths}  \hfill & \hbox{number of deaths} \hfill &
\hbox{number of deaths} \hfill \cr
\qtb
\hbox{}  \hfill & \hbox{per week}  \hfill & \hbox{per week} \hfill &
\hbox{per week} \hfill\cr
\noalign{\hrule}
\qth 
\hbox{Week 1}  \hfill & 4.00\quad [1.00] &  4.00\quad [1.00] & 13.\quad [1.00] \cr
\hbox{Weeks 2,3,4}  \hfill & 1.70\quad [0.42] &  2.33\quad [0.57] & 8.3\quad [0.64]\cr
\hbox{Weeks 5-12}  \hfill & 0.87\quad [0.22] &  1.63\quad [0.40] & 6.2\quad [0.47]\cr
\qtb
\hbox{Weeks 13-24}  \hfill & 0.54\quad [0.13] &  0.66\quad [0.16] & 5.9\quad [0.45]\cr
\noalign{\hrule}
} $$

\vskip 1.5mm
Notes: The number within brackets show the data in normalized form 
(first week=1). 
At first sight one may be tempted to interpret the decrease
in the number of suicides per week as reflecting a transient state
marked by a reorganization of social ties. However, the parallel observations
for deaths by natural causes and by accidents clearly require different
explanations. In the case of death by natural causes a key-factor is the
fact that terminally ill prisoners are released so they can die at home or
in hospital. Alcohol is also an important factor. 
\qL
Source: Sattar (2001, p. 34)
\vskip 2mm

\hrule
\vskip 0.5mm
\hrule

\normalsize

\end{table}


The fact that some caution must be exercised is suggested by the second
column in table 1a. It shows that for accidental deaths too,
there is also a surge in the first week after
release. In contrast to suicide however the rate remains almost 5 times
higher than in prison even in subsequent weeks. This is not really
surprising for in prison the risk of being run over by a car or to freeze to
death while sleeping in the street is zero. 
Table 1b provides additional information about the decrease in
standard mortality rates in the weeks after release. We see that the pattern
is basically the same either for suicide, for deaths by natural causes or
for deaths by accident. The following precisions may be useful for the
interpretation of these data.
\qbu Prisoners afflicted with a fatal illness are often released  in their
last weeks of life. This policy may well account for the fact that the standard
mortality rate for overall deaths is about 3 times higher for ex-prisoners
than for prisoners. It may also account for the surge of natural deaths
in the first week after release if the policy is to release prisoners in the
very last stage of their illness. To make that point clearer one would need
to know the criteria on which release is decided.
\qbu While the availability of alcohol is fairly limited in prison, all of
a sudden it becomes available once a prisoner is released. In a general
way, one knows that there is a strong link between alcohol consumption
and suicide rates. This effect is documented convincingly by Nizard 
(1998, p. 2). Naturally, it can be argued that
the excessive consumption of alcohol is in itself a kind of suicidal
behavior. The point that we made for alcohol can also be made for
drugs.
\qbu The ex-prisoners considered in Sattar's study 
on which the present discussion is based were under 
community supervision. In fact, it is for
this very reason that their death could be recorded.
However, when the period of supervision of an ex-prisoner comes
to an end he leaves the sample and his death will  no longer be recorded.
If the period of supervision of an ex-prisoner lasts only 4 weeks
his death will not be recorded even if it occurs only 5 weeks after
release. Naturally, this procedure tends to inflate the number of deaths
which occur shortly after release. To some extent this may explain
the decrease pattern observed in the three columns of table 1b. Unfortunately,
this effect cannot be controlled for.
\qpar

{\bf Transition from civil to military status}\quad
Being enlisted represents a major change in social environment
and requires a reorganization of social ties. One would therefore expect
a surge in suicide rate followed by subsequent decrease just as observed
in Fig. 1 after incarceration. In a thesis published in France 
almost a century ago,
Botte (1911) provides a comparison between
the suicide rate of soldiers in their first army year 
(we call them ``new'') and soldiers
who had been in the army for more than one year (we call them ``old'').
Remember that 
at the time military service lasted 3 years for French draftees. On average,
for the 1888-1907 period for which Botte provides data, the suicide rates
for new and old soldiers were 
$ 26.0/10^5 $ and $ 20.6/10^5 $ respectively.
The ratio ``new''/``old'' is 1.26. As expected this ratio is indeed larger
than one. However, it is markedly smaller than for the shift to prison life.
From the data given in {\it Bridge} it results that for jail inmates
(i.e. for less than one year) the average suicide rate is $ 130/10^5 $
 (over the period 1981-1987) while it is $ 21/10^5 $ for inmates
in prison for more than a year). In this case, the ratio is: $ 130/21=6.2 $.
\qpar

Botte also gives another comparison which can be of interest
in connection with our subsequent discussion of family ties.
He shows that over the period 1878-1908,
the average suicide rate was twice as high for
troops stationed in the French colony of
Algeria as for troops stationed in France. 
A similar result is given in another study for the earlier period 1865-1869: 
the average suicide rate was 1.28 times larger for troops in Algeria
than for the total army (Cristau 1874). The troops in Algeria differed
from those in France in two main respects: (i) They were separated
from their relatives to a greater extent than troops stationed in France.
(ii) They comprised indigenous soldiers. Both factors may
have contributed to the higher rate; the first factor can be connected
to our subsequent discussion regarding the impact of family links.

\qA{Open vs. closed prisons}
Open prisons are without walls or bars and offer the possibility to work
or study outside of prison during working hours, the evenings and
nights being spent in the prison. 
Open prisons are for prisoners who are believed not to be a risk to the
public. For instance, total objectors who refuse all service in the 
conscription system including non-military service may be assigned
to an open prison. The data on which we rely in this paragraph come
from two detailed reports published by the British Home Office
(Prison Statistics 2001, HPMS Annual Report 2004). 
\qpar

Table 2 shows that suicide rates are at least 10 times higher in closed 
than in open institutions. The decision to affect a prisoner to an open rather
than to a closed institution is taken in the days or weeks following the
incarceration. In other words, the initial trauma caused by the incarceration
is the same for all prisoners whatever institutions they belong to 
subsequently. In short, this observation seems to disprove the trauma
explanation. On the contrary, it has a natural explanation in the tie
perspective in so far as open prisons are of course more conducive
to the establishment of social ties. One could be tempted to say that the
two populations are not the same which is quite true. However, it
is by no means obvious that a prisoner who represents a threat to the public
(and for that reason will be affected to a closed prison), necessarily has
a higher ``propensity'' for committing suicide. 
\qpar


\begin{table}[htb]

 \small 

\centerline{\bf Table 2\quad Comparison of suicide rates in open versus
closed prisons, 1994-2001}

\vskip 3mm
\hrule
\vskip 0.5mm
\hrule
\vskip 2mm

$$ \matrix{
\tvi 
 &\hbox{Category of prisoners}  \hfill & \hbox{Average population}  \hfill & 
\hbox{Average number } \hfill & \hbox{Annual} \hfill \cr
 &\hbox{}  \hfill & \hbox{of prisoners}  \hfill & 
 \hbox{of suicides} \hfill & \hbox{suicide rate} \hfill \cr
\qtb
 &\hbox{}  \hfill & \hbox{}  \hfill & 
\hbox{per year} \hfill & \hbox{per 100,000} \hfill \cr
\noalign{\hrule}
\qth 
 &\hbox{\bf Young prisoners (15-21)} \hfill & & & \cr
 1 &\hbox{Open Young Offender Institution} \hfill & \phantom{4,}380 & 0& 
\phantom{13}0\cr
2 &\hbox{Closed Young Offender Institution} \hfill & 4,690 & 6 & 138\cr
 &\hbox{} & & & \cr
 &\hbox{\bf Adult prisoners (over 21)} \hfill & & & \cr
 3 &\hbox{Open Male Training Institution} \hfill & \phantom{2}3,430 & 0.12 & 
\phantom{4}3.6\cr
\qtb
 4 &\hbox{Closed Male Training Institution} \hfill & 19,190 & 11.1 & 47.4 \cr
\noalign{\hrule}
} $$

\vskip 1.5mm
Notes: 
There is a difference of at least a factor 10 for suicide rates in open versus
closed institutions.
Are the populations in the different categories large enough to
lead to substantial suicide rates. Under the assumption of a suicide rate
of about $ 15/10^5 $ it would require a sample of more than 
$ 10^5/(15\times 8) = 830 $ prisoners to get at least one suicide per year.
On this basis it can be seen that all samples are large enough except 
possibly the first one. In the later case, over the 8 years one would expect
$ 380\times 8 \times (15/10^5) = 0.46 $ deaths. The observation gave 0, but
it could have been 1, 2 or 3 without altering our conclusion. Note that
the closed male institutions belong to the
categories B and C defined in the statistics of the Home Office.
\qL
Sources: Prison Statistics, England and Wales 2001, HPMS Annual Report
2004.
\vskip 2mm

\hrule
\vskip 0.5mm
\hrule

\normalsize

\end{table}

In the next section we turn to another situation which results
in the severance of social ties, namely emigration.

\qI{Suicide rates of immigrants}

In the previous section we considered the weakening of social ties when
a person is put into custody. Naturally, this is only one of many possible
mechanisms leading to a weakening of social ties. The process of emigration
is another obvious candidate. Back in the 19th century, when an individual
or a family emigrated from, say, Italy to the United States, it meant an
almost total severance of the contacts with relatives, friends and neighbors
left behind in Italy. Furthermore, until the language barrier was surmounted, it
was not easy to establish new social ties in the United States except of
course with other Italian immigrants. As a matter of fact, this is certainly
the main factor which made Italian immigrants live together in areas
such as ``Little Italy''  in South Manhattan. 
\qpar

The tie perspective would make us expect suicide rates of 
Italian immigrants in
New York City  to be higher; but higher than what? Higher than
the average suicide rate in New York or higher than the average suicide rate
in their region of origin? Fig. 2 answers this question. 
  \begin{figure}[htb]
   \centerline{\psfig{width=15cm,figure=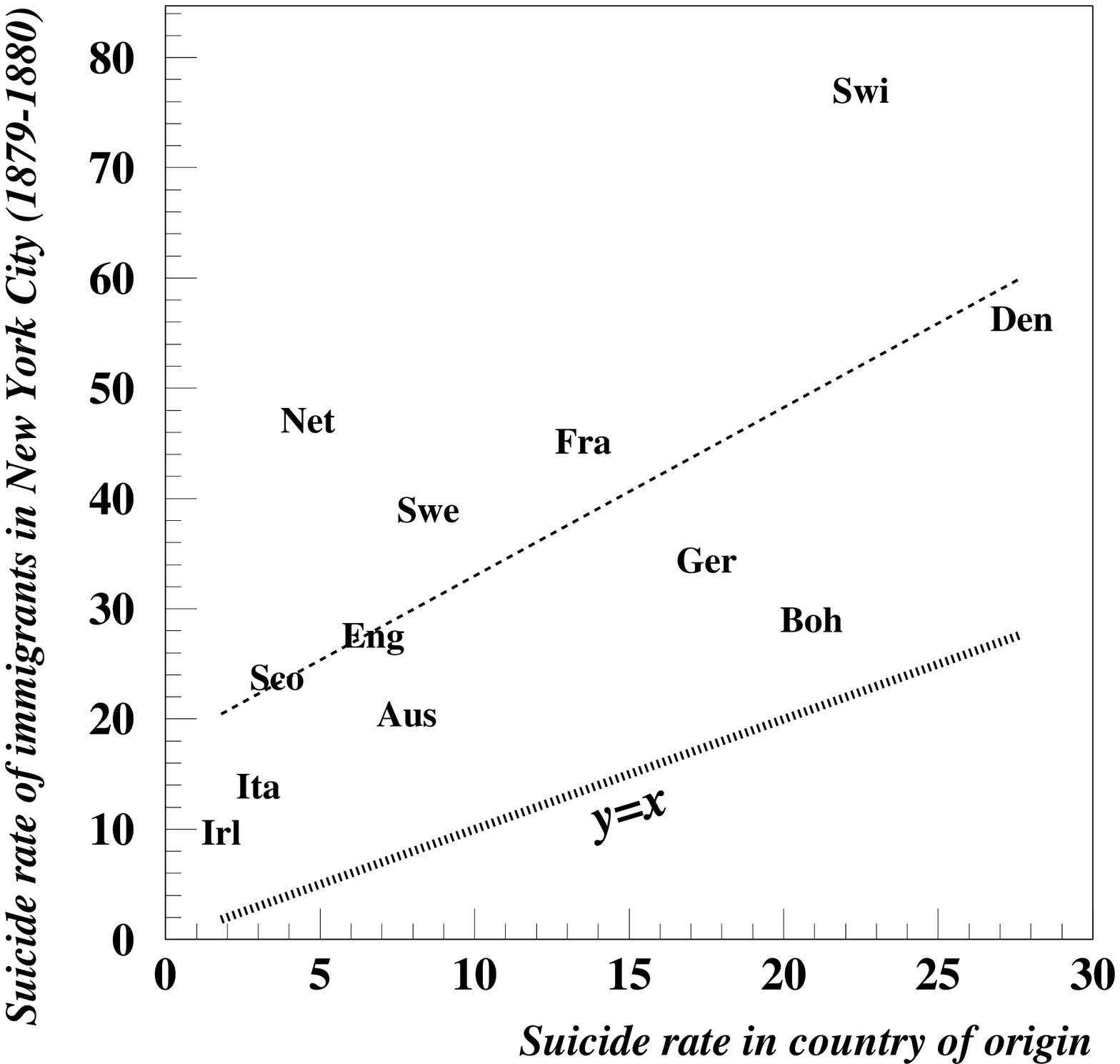}}
    {\bf Fig.2: Suicide rates of immigrants in New York City (1879-1880)}.
{\small Each abbreviation corresponds to immigrants from a different
country of origin (Boh=Bohemia, Sco=Scotland, the others are fairly
transparent). It can be seen that the suicide rates in New York City are
higher than in the region of origin which is consistent with the
rearrangement of social ties brought about by the process of emigration.
There is a marked correlation
between the levels recorded in the country of origin and in New York 
City. This consistency
provides an overall reliability check of suicide statistics. 
The correlation is equal to 0.70 (with a confidence
interval at probability 0.95 extending from 0.21 to 0.91) and the slope of
the regression line is $ 1.5 \pm 1 $. The line $ y=x $ is shown for the purpose
of comparison}.
{\small \it Sources: Durkheim (1897), Krose (1906), Nagle (1882)}.
 \end{figure}
It turns out that
there is a high correlation between the suicide rates of immigrants in
New York and the suicide rates prevailing in their country of origin. 
In other words, suicide rates seem to display a great inertia. 
For groups of immigrants their suicide rate seems to be a cultural 
attribute just like their language or religion. We do not yet 
have a comprehensive
explanation of this effect but in a sense it is not
completely surprising. For instance, the fact that suicide is strongly prohibited
by the Catholic Church will tend to lower the impact of
suicide among Catholics both in the country of origin and in New York.
The difficulty is that this can be understood
in a number of different ways. (i) It can induce families and
physicians (at least if they are Catholic too) to fake the evidence. In this
case, it would be nothing but a statistical bias%
\qfoot{This kind of arguments was developed by Douglas (1967) who
expressed great doubts about the validity of suicide statistics. 
That there are indeed measurement problems can hardly be denied.
For instance, in 1987 the official suicide rate in 
Egypt was equal to $ 0.1/10^5 $
for males and to $ 0.0/10^5 $ for females (Schmidtke et al. 1999).
However,
the above correlation shows that if this effect exists,
it is only of marginal importance for European countries.
On average, the rates of suicide recorded in New York
(mostly by American physicians one can suppose) are consistent with
the rates recorded in the countries of origin which shows that the
latter are on the whole trustworthy.}%
.
(ii) Alternatively, one can also suppose that suicide rates are really lower
among Catholics. Before trying to find an explanation one would have
to check whether this is really true everywhere or also depends upon 
broad regional traditions. We plan to study this question
more carefully in a subsequent paper.
\qpar

Let us now come back to the question of the two interpretations.
Fig. 2  shows that for each country of origin the suicide rate in New York
is increased, the increment being of the order ot 18 per 100,000. 
This increment has a natural interpretation in the tie perspective.
At first sight, one may think that it is also consistent with the
trauma hypothesis. It cannot be denied that the act of emigrating
represents a cultural and psychological shock. However, there is
a problem with the way this shock subsides in the course of time.
In the case of an incarceration we have seen that suicide rates
revert to their pre-incarceration level within two years approximately.
In the case of Fig. 2 we do not know how long these immigrants had
been in New York, but it is probably safe to assume that on average
they had been in New York for more than two years. It would be difficult
to understand then why the emigration shock (presumably less traumatic
than an incarceration) would last several years. On the contrary, a 
transient period of several years for the formation of new links is
in agreement with the time it takes to learn English and to make new
acquaintances.
\qpar

To our best knowledge there have been very few studies documenting
suicide rates among immigrants. The topic which we consider
in the next section, namely the influence of family links, has given rise
to a larger number of studies.

\qI{Influence of family ties}

One of the best discriminating tests between the trauma and tie models
is to observe the effect of changing family ties. 
Table 3 summarizes the evidence for four different countries and for
time intervals which cover more than one century. On average, the 
suicide rate for unmarried men is 2.4 times higher than for married men;
for women the ratio is 1.9. We deliberately excluded the evidence for
separated, divorced or widowed individuals. For these categories
the suicide rates are also higher than for married people but they
imply a traumatic event which could serve as an 
alternative explanation. On the contrary, for people who never married
one can hardly invoke a possible trauma. 
\qpar


\begin{table}[htb]

 \small 

\centerline{\bf Table 3\quad Influence of family ties on suicide rates}

\vskip 3mm
\hrule
\vskip 0.5mm
\hrule
\vskip 2mm

$$ \matrix{
\tvi 
 &\hbox{Time interval}  \hfill & \hbox{Country}  \hfill & 
\hfill \hbox{Males} \hfill & \hfill \hbox{Females} \hfill \cr
 &\hbox{}  \hfill & \hbox{}  \hfill & 
 \hbox{ratio of } \hfill & \hbox{ratio of} \hfill \cr
 &\hbox{}  \hfill & \hbox{}  \hfill & 
 \hbox{suicide rates:} \hfill & \hbox{suicide rates:} \hfill \cr
\qtb
 &\hbox{}  \hfill & \hbox{}  \hfill & 
 \hbox{not married / married} \hfill & \hbox{not married / married} \hfill \cr
\noalign{\hrule}
\qth 
1 & 1889-1891 & \hbox{France}\hfill& 2.80\pm 0.23 & 1.56\pm 0.43\cr
2 & 1881-1890 & \hbox{Switzerland}\hfill& 1.66\pm 0.24& 1.34\pm 0.41\cr
3 & 1968-1978 & \hbox{France}\hfill& 2.69\pm 0.66& 2.24\pm 1.10\cr
4 & 1981-1993 & \hbox{France}\hfill& 2.34\pm 0.33& 2.15\pm 0.91\cr
5 & 1990-1992 & \hbox{Queensland}\hfill& 2.67\pm 1.20 & 2.11\pm 1.60\cr
6 & 1998 & \hbox{Australia}\hfill& 2.21\phantom{\ \pm 1.20} & 
2.00\phantom{\ \pm 1.60} \cr
7 & 1970-1985 & \hbox{Norway}\hfill &  & 1.78 \pm 0.24\cr
 &  & &  & \cr
\qtb
 &  & \hbox{\bf Average}\hfill &  2.40 \pm 0.24 & 1.88\pm 0.32\cr
\noalign{\hrule}
} $$

\vskip 1.5mm
Notes: In most cases (except 6) detailed data by age interval were
available. We computed the ratio of the suicide rates in each age 
interval and then the average $ m $ and the standard deviation $ \sigma $
of the ratios.
The results in the table are given in the form $ m \pm \sigma $.
It should be noticed that the category ``married'' includes both
``married without children'' and ``married with children''; consequently,
the reduced suicide rates for ``married'' should not be attributed solely
to marriage but to the combined effect of being married and having children.
\qL
Sources: 1: Durkheim (1897); 2: Halbwachs (1930); 3,4: Besnard (1997);
5: Cantor et al. (1995); 6: Steenkamp et  al. (2000); 7: H\o yer et al. (1993).
\vskip 2mm

\hrule
\vskip 0.5mm
\hrule

\normalsize

\end{table}


Furthermore, there is evidence in the literature showing that married
people with children have smaller suicide rates than married people
without children. Unfortunately, most of these observations are indirect
because death certificates usually do not indicate the number of children
of the deceased. As a result these data require careful discussions
which we leave for a subsequent study.
\qpar

To sum up, one can say that the tie perspective provides a natural
interpretation for suicide rate differentials in various family situations.
However, there is one observation which, at first sight, it is not
able to accommodate, namely the differential between males and females.
In European countries and in the United States%
\qfoot{It does not seem to be true in China or in some parts of India}%
, 
whatever their respective family situations, males
have suicide rates which are 2 to 3 times higher than females in same
family situations. If one tries to interpret this effect in the tie model
one is led to assume that females have stronger family ties than
males. While this makes sense for married women with children (especially
if they are housewives) it does hardly explain why it is also true for
unmarried women respective to unmarried men. In that case one would
have to assume stronger links with parents, brothers, sisters and other
close relatives. 
\qpar

It is possible to perform a plausibility test of the former argument. Over
the past 60 years, as more and more women have taken up a job
their links with their family tended to become similar
to those of working men. If the tie model holds one would therefore
expect a narrowing of the suicide rate gap. This is indeed what one
observes as shown in table 4. 



\begin{table}[htb]

 \small 
\centerline{\bf Table 4 \ Suicide rates of unmarried people: male vs. female}
\vskip 3mm
\hrule
\vskip 0.5mm
\hrule
\vskip 2mm

$$ \matrix{
\tvi 
\hbox{} \hfill & 1889-1891 & 1968-1973 & 1979-1983 & 1989-1993\cr
\noalign{\hrule}
\qth 
\hbox{Unmarried males, 30-59 y.} \hfill & 101 & 65.4 & 76.5 & 65.8 \cr
\hbox{Unmarried females, 30-59 y.} \hfill & 16.7 & 17.2 & 24.0 & 23.1 \cr
\qtb
\hbox{Ratio M/F} & 6.0 & 3.8 & 3.2 & 2.8 \cr
\noalign{\hrule}
} $$
\vskip 1.5mm
Notes: For married males versus married females, the 
decrease in the ratio of suicide rates was smaller: from 3.20 in
1889-1891 to 2.45 in 1989-1991 (for the same 30-59 year interval).
\qL
Source: Besnard (1997).
\vskip 2mm

\hrule
\vskip 0.5mm
\hrule

\normalsize

\end{table}


\qI{Effect of global traumas}

The expression ``global traumas'' refers to dramatic event which affect
the society as a whole. The
outbreak of a war or a major disaster belong to this class of events.
It has been claimed that such events reduce the suicide rate
because, allegedly, of the polarization they create in human minds.
If true, this effect would be a  strong argument in favor of
the trauma model. In the tie model one would not expect
any effect based on purely psychological factors. 
In the case of a war lasting several years, the expectation must
take into account the rearrangement of family ties occasioned by
the war conditions (e.g. the fact that many husbands are enlisted).
\qpar

In this section we make the following points.
\qbu Short events which do no imply
a reorganization of social ties do {\it not}
have any incidence on suicide rates. We  consider two episodes
of this kind (i) The hyperinflation in Germany in 1923 (Fig. 3a)
(ii) September 11, 2001 (Fig. 3b).
\qbu Wars lasting several years have no clear-cut influence on the
suicide rate of females. As far as males are concerned, there is a decrease
in suicide rates. This decrease is difficult to interpret as
we explain below. We consider 3 episodes of this kind
(i,ii) World War I  in Germany and in
the United States (Fig. 3a and 3c)
(iii) World War II in the United States (Fig. 3c)
(iv) The Korean War in the United States (Fig. 3c).
\qbu In addition we also consider the period of the Great Depression in
the United States. This an episode which is of a somewhat different
kind than the previous ones. It is marked by a substantial increase in
the suicide rates of males but the suicide rate of females increase only
slightly (Fig. 3c).
\qpar
  \begin{figure}[htb]
   \centerline{\psfig{width=15cm,figure=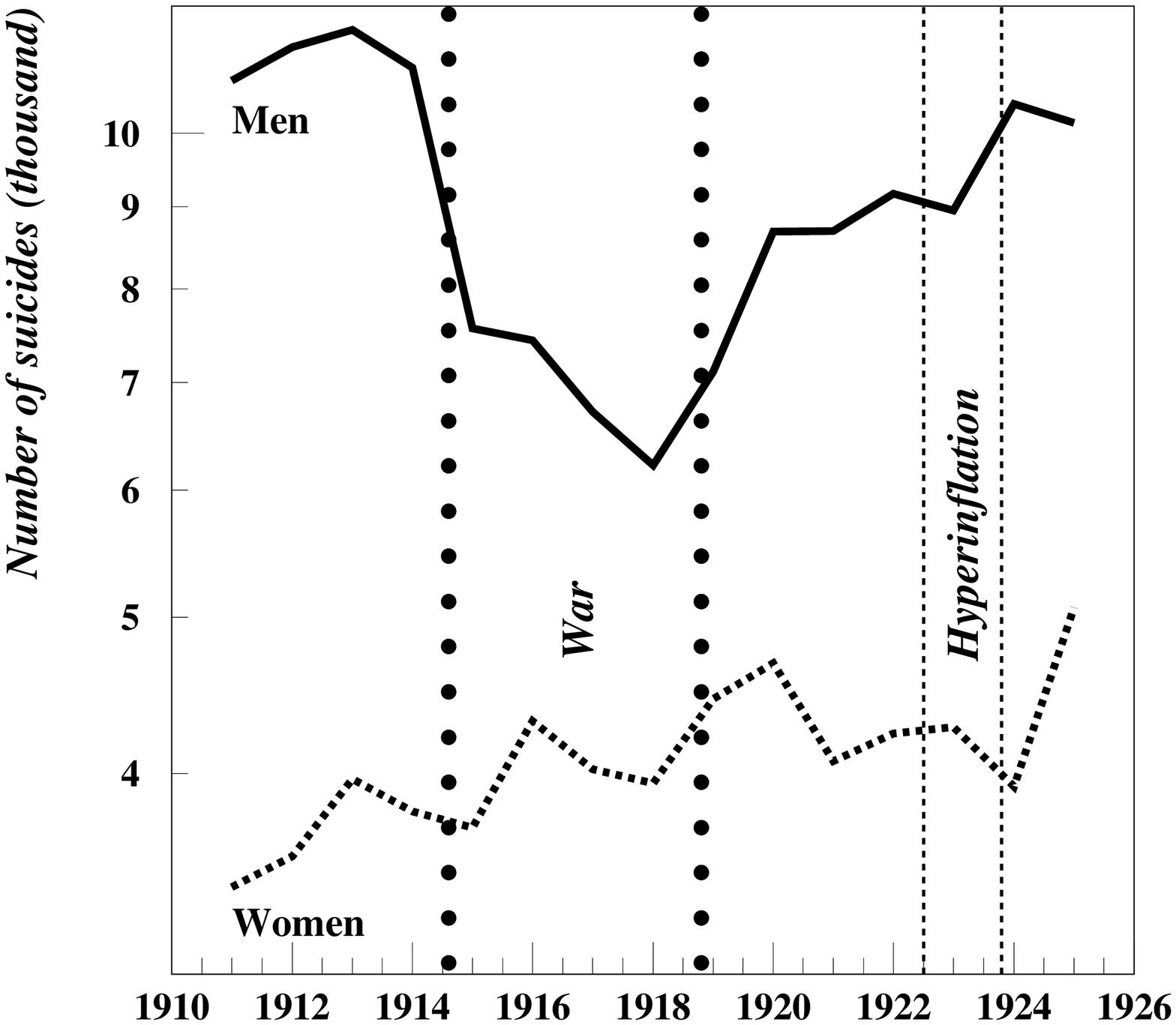}}
    {\bf Fig.3a: Number of suicides in Germany 1911-1924}.
{\small The fact that the direction of variation is not the same
for males and females during the war and the hyperinflation episode
shows that even major events have little effect on 
suicide rates}.
{\small \it Source: Statistisches Jahrbuch for das deutsche Reich 
(various years)}.
 \end{figure}
  \begin{figure}[htb]
   \centerline{\psfig{width=15cm,figure=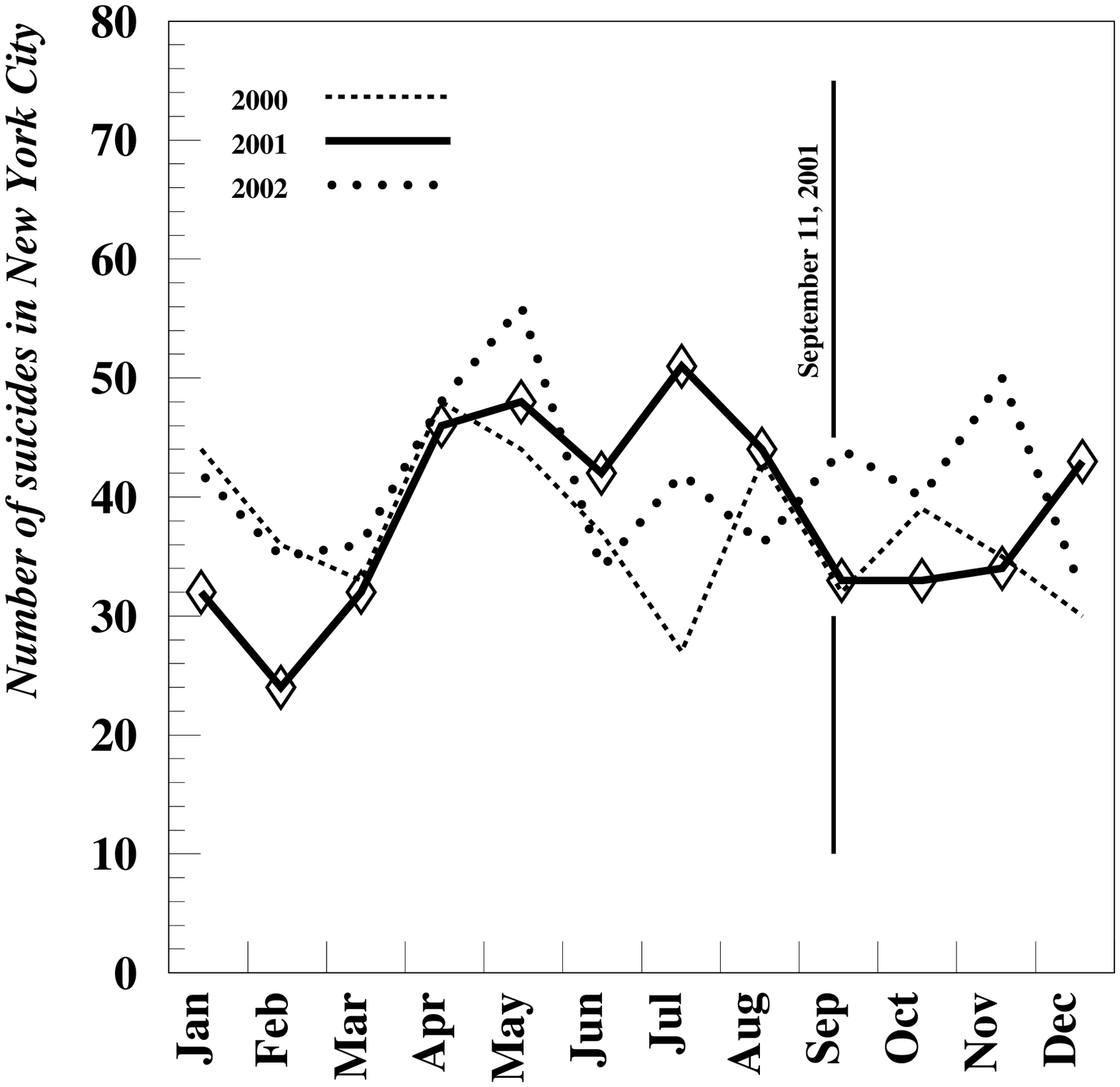}}
    {\bf Fig.3b: Number of suicides in New York City  before and after
September 11, 2001}.
{\small The graphic shows the number of suicides in each month of
the years 2000, 2001 and 2002. The data for 2000 and 2002 are
for the purpose of comparison. It can be seen that between August and
December the curves for 2000 and 2001 are very close which indicates
that 9/11 has had no significant impact on suicide rates}.
{\small \it Source: I would like to express all my gratitude to
Dr. Joseph Kennedy of the New York City Department of Health for sending
me these data}.
 \end{figure}
  \begin{figure}[htb]
   \centerline{\psfig{width=15cm,figure=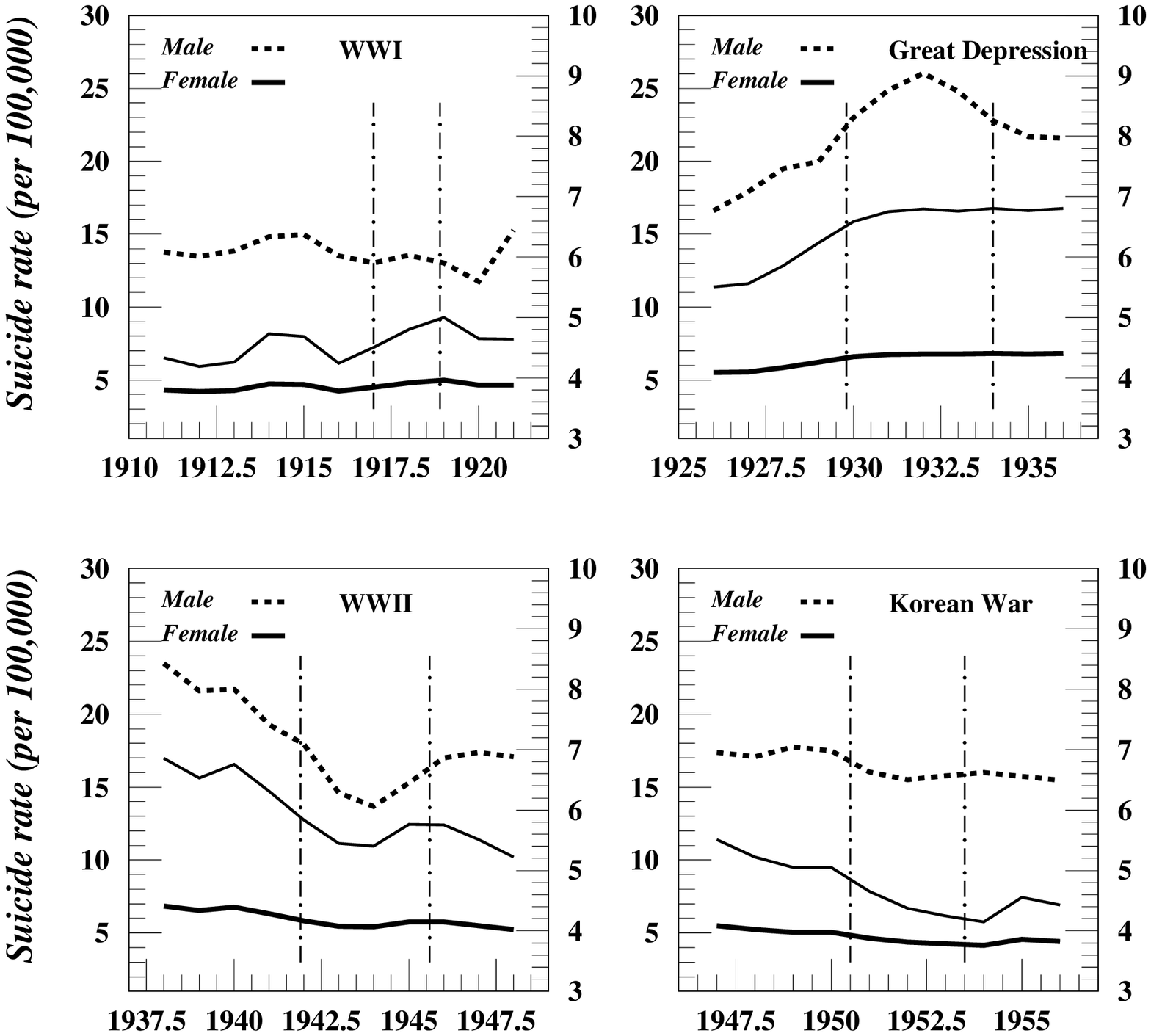}}
    {\bf Fig.3c: Suicide rates in the United States during global shocks}.
{\small The broken lines and the thick solid lines give the
suicide rate for males and females respectively (left-hand scale);
the thin solid lines provide a magnification of the fluctuations
of female rates (right-hand scale). The Great Depression is the only
case where the female and male rates move in the same direction. For the
three other cases, suicide rates do not respond in a consistent and
significant way}.
{\small \it Source: Historical Statistics of the United States (1975).
Prior to 1930, the data must be handled with caution because several
states (especially in the West) do not belong to the Registration Area.
The dates of entry of each state into the Registration Area can be found
in Candiotti et al. (1948)}.
 \end{figure}
Next we discuss these episodes in more detail.
\qpar

{\bf Hyperinflation of 1923 in Germany}\quad It should be recalled that
this inflationary episode suppressed the huge internal war debt
of the German government. The hyperinflation ended in November 1923
when a new mark (the so-called {\it Rentenmark})
was introduced on the basis of 
$ 1\ \hbox{new mark} = 10^{12}\ \hbox{old mark} = 0.24\ \hbox{US dollar} $.
Needless to say,
at the same time as it extinguished the debt the hyperinflation also
obliterated all savings which had not been exchanged for dollars.
In other words, it was really a traumatic event for a large part of the
population. Yet, Fig. 3a shows that there is no clear-cut effect on
suicide rates: the suicide rate of males increases (but it may well
have been the continuation of a trend which began in 1918) while the
suicide rate of women shows a small decrease. 
\qpar

{\bf September 11, 2001}\quad We selected the data
in a way  aimed at magnifying any possible effect, even a small one.
(i) Instead of looking at suicide rates for the U.S. or for
New York State, we focused on data for New York City where the impact
of September 11 would be expected to be largest. 
(ii) Instead of annual or quarterly
data, we analyzed monthly data so as to be able to identify even
an effect of short duration. 
Nonetheless, Fig. 3b shows that there is no perceptible effect
\qfoot{It would be meaningless to perform a test of significance.
Such a test would rely on assumptions regarding the statistical distribution
of suicides (in particular its standard deviation) 
and would not in this case be more reliable than visual inspection.
We see that over September-December, the 2001 curve 
is very close to the 2000 curve; on the contrary the 2002 curve deviates
from the two others in spite of the fact that there was no major shock in
the fall of 2002.}
.
\qpar

{\bf Wars}\quad Two preliminary remarks are in order.
(i) One should concentrate on countries whose territory was not
occupied. Why? Consider the
case of France. During the duration of Word War I the northern part
of its territory was occupied by Germany. As a result the suicide statistics
could not be collected and 
recorded by the French statistical agency. A similar observation
applies to World War II because the departments of the north and east
came under direct German rule and were no longer considered as belonging
to France.
(ii) It is not at all obvious that suicides occurring 
in the army in times of war are truthfully recorded.
Why? in a general way committing suicide is seen
negatively both by the
military and the families. This is far more true
in times of war. For field commanders
suicides are seen as a form of desertion. As a result,
it is likely that many suicides are declared as being accidents.
``Unfortunately, he shot himself while cleaning his riffle''.
Unless one has positive proof to the contrary, suicide rates among enlisted
men should therefore not be considered as reliable, especially
if they are low. 
For these two reasons one must be particularly careful in 
selecting the evidence%
\qfoot{In Chesnais (1976, p. 56,61) there are two graphics comparing
suicide rate changes in several
countries (Britain, France, Germany, Sweden 
Switzerland, USA) 
during World War I and II. As many of 
the curves show a marked dip, the evidence at first seems
quite impressive. In both wars the dips are particularly large in France.
However, as France cumulates the two causes of bias
these dips are largely meaningless. In World War I 
Switzerland has a big trough. This would be a good argument in favor
of a purely psychological effect for in this case none of the above
bias can be invoked. However, in World War II 
Switzerland does not display any trough whatsoever.}%
.
\qpar

{\bf Effect of World War I in Germany}\quad 
Fig. 3a shows that between 1914 and 1918
there was a 10\% {\it increase} in the suicide rate of women. 
There is therefore no ground for a psychological effect due to the war
\qfoot{There is a 50\% decrease in the suicide rate of men. Its magnitude
remains an open question for the trough is too big to be 
be explained away by unrecorded suicides among the military.}%
.
\qpar

{\bf Effect of World War I in the United States}\quad 
As in Germany we observe an {\it increase} in the suicide rate of women,
(albeit of small magnitude). 
In the tie model these increases can be attributed to the severance 
(temporarily of permanently) of the link between husbands and wives.
This interpretation could be tested against suicide rate data by age.
\qpar

{\bf Effect of World War II and of the Korean War in the United States}\quad 
Once again we have a similar pattern: small increase in the suicide
rate of females (if one leaves apart the initial drop which continued
the previous trend) and a drop in the suicide rate of men.
\qpar


{\bf Effect of the Great Depression in the United States}\quad The Great
Depression was much more than a psychological trauma; it deeply
affected the life of millions of people. Therefore it is hardly surprising
that it has had a substantial influence on suicide rates. In this case
we see that the rates of both males and females display parallel evolutions.

\qI{Conclusion}
We have collected, selected and compared data sets in order to
discriminate between the tie model and the trauma model.
It turns out that there is a broad array of evidence in favor of the tie model.
Fig. 4 schematically summarizes this model.
  \begin{figure}[htb]
   \centerline{\psfig{width=8cm,figure=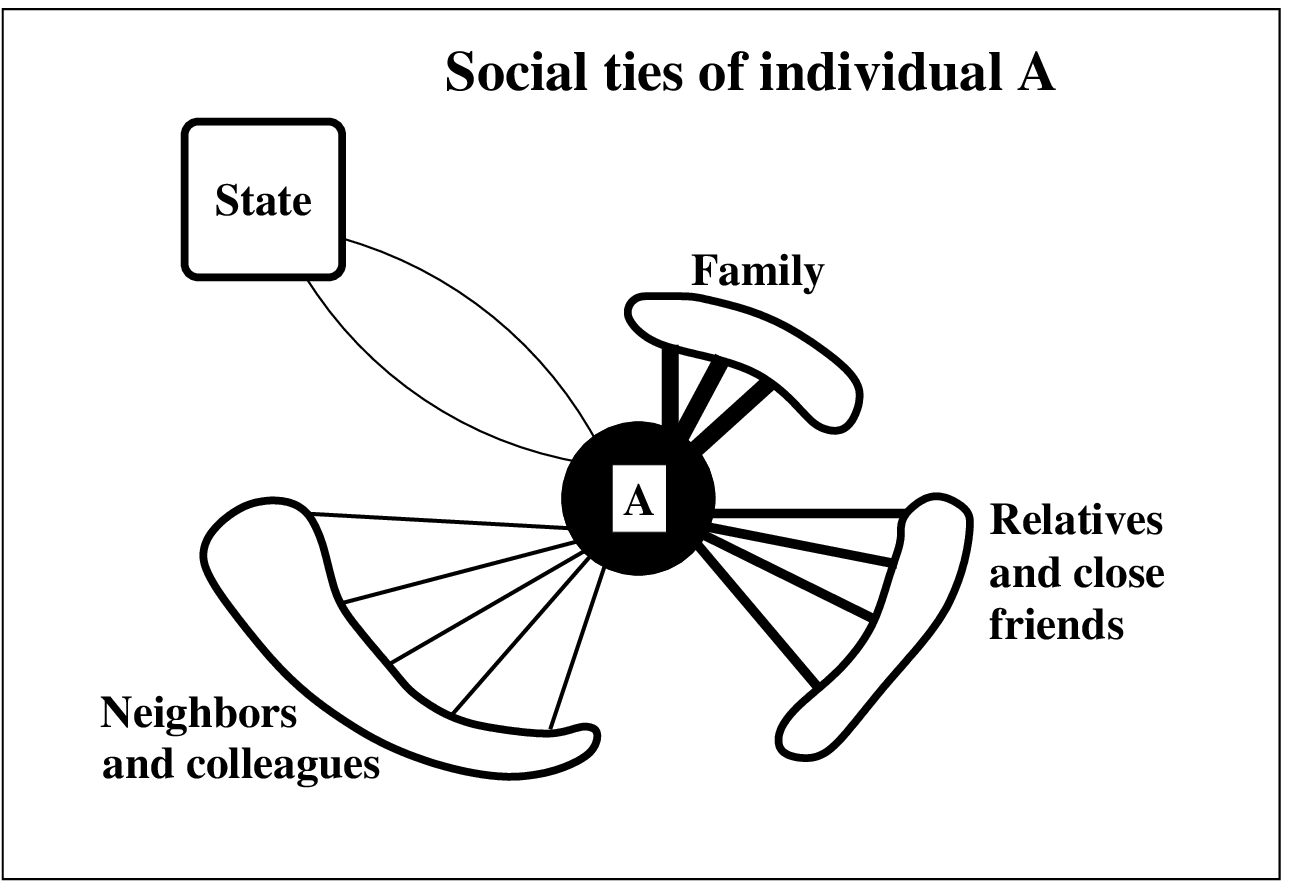}}
\vskip 3mm
    \centerline{\bf Fig.4a: Schematic representation of interpersonal ties.}
{\small The box labeled ``State'' symbolizes the society considered
as a global entity.}
 \end{figure}
  \begin{figure}[htb]
   \centerline{\psfig{width=8cm,figure=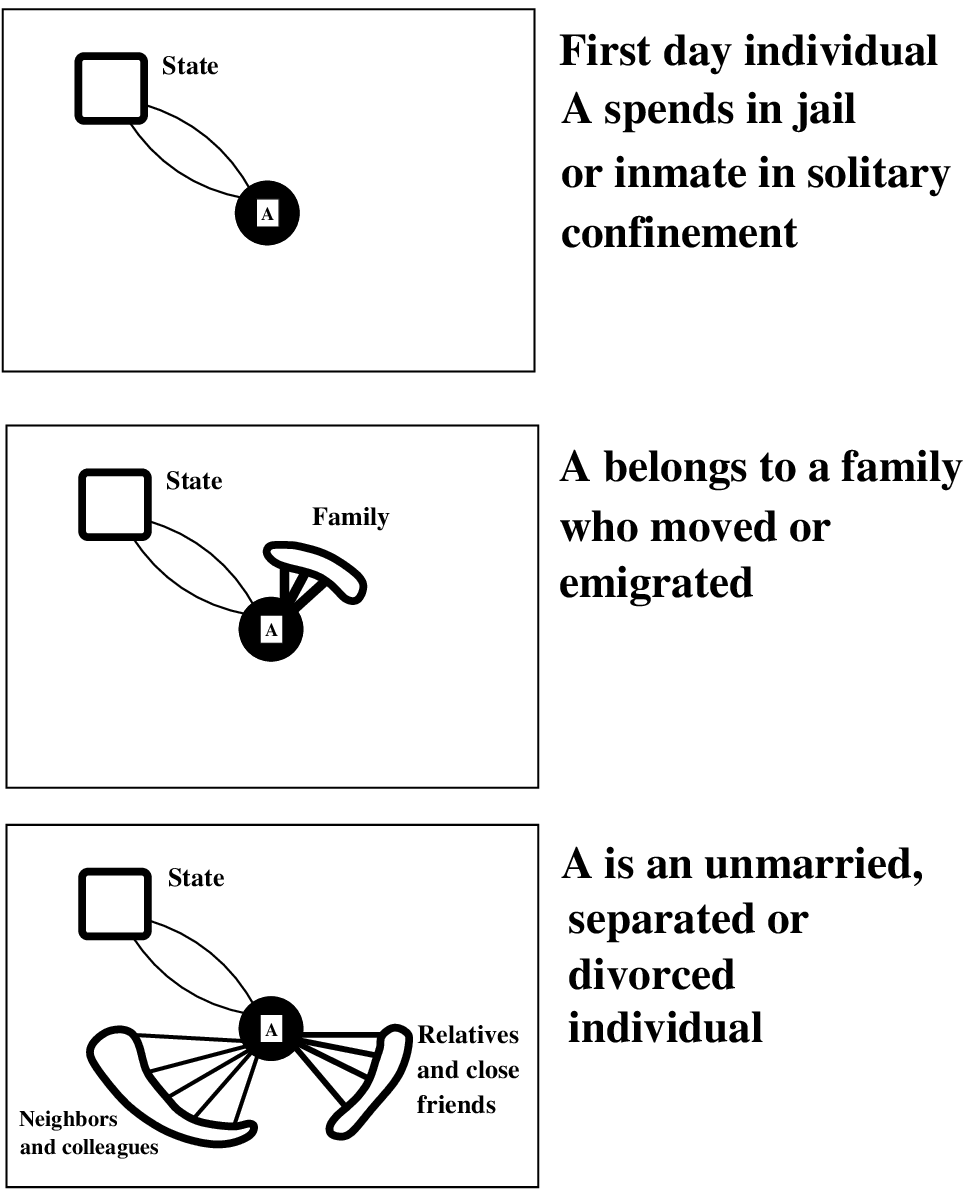}}
\vskip 3mm
    \centerline{\bf Fig.4b: Schematic representation of the social ties in
three situations considered in the paper.}
 \end{figure}

What remains to be done? To see this point more clearly it may
be helpful to physicists to 
draw on a comparison with physics. Suppose we want to
study the movements of 
a number of pendulums which differ in their initial velocities as well
as in their shapes. This will be a very difficult task
because the movements are superpositions of (at least) three
different effects: movement of a spherical pendulum, movement
of a pendulum whose mass is not a sphere, movement of a Foucault 
pendulum. It is only by distinguishing and sorting out these different
mechanisms that it will become possible to get  insight and understanding.
The situation is similar with suicides. 
In this paper we studied the effect of severing ties. This gave rise
to transient and permanent effects. 
Furthermore there are several side effects which must be ``filtered out'';
one can for instance mention the following: (i) The means and agencies
for self-harm available to a potential suicidee greatly affect the suicide 
rates. (ii) The availability of drugs, whether medical drugs, narcotics or
alcohol, can have a marked effect on suicide rates. 
\qpar

A last word is in order. Often econophysicists are blamed by social 
scientists for not giving full credit to former studies. Did we commit
the same fault in this paper? It is true that the phenomenon of suicide
has been studied at least since 1850%
\qfoot{In {\it Bridge} the reader will find a selected bibliography which
goes back to the mid-nineteenth century.}%
.
However, very few studies have been conducted in the comparative
and analytical spirit pioneered by Emile Durkheim.
Even among the later, 
one finds many comparisons which have not been conducted carefully 
enough (an example was given above for the effect of wars on suicides)
Instead of bringing more
light such comparisons just add to the confusion. Physicists can
rely on a century-long tradition of eliminating spurious effects,
chasing up main determinants and
disentangling complicated phenomena. Accordingly, 
econophysicists should be
in a good position to throw some new light on this question.

{\bf Acknowledgements}\quad I must most grateful to Professors
Bernard Diu and Olivier G\'erard for their helpful comments.

\vfill \eject

{\large \bf References}
\vskip 5mm

\qparr
Besnard (P.) 1997: Mariage et suicide: la th\'eorie durkheimienne de la 
r\'egulation conjugale \`a l'\'epreuve d'un si\`ecle. 
Revue Fran\c{c}aise de Sociologie 38,735-758.

\qparr
Blom-Cooper (L.) 1987: The penalty of imprisonment.
Tanner Lectures on Human Values. 
Delivered at Clare Hall, Cambridge University, 
November 30 - December 2, 1987. 

\qparr
Bonanno (G.), Lillo (F.), Mantegna (R.) 2001: Levels of complexity
in financial markets. Physica A 299,16-27.

\qparr
Botte (G.) 1911: Le suicide dans l'arm\'ee. Etude statistique, \'etiologique,
et prophylactique. Thesis. Lyon. 

\qparr
Candiotti (C.), D\'erobert (L.), Moine (C.), Moine (M.) 1948: Consid\'erations
statistiques sur le suicide en France et \`a l'\'etranger.
Institut National d'Hygi\`ene. 

\qparr
Cantor (C.H.), Slater (P.J.) 1995: Marital breakdown, parenthood, and suicide.
Journal of Family Studies 1,2,91-102.

\qparr
Central Bureau of Statistics of Norway 1978: Historical statistics.
Oslo. 

\qparr
Chesnais (J.-C.) 1976: Les morts violentes en France depuis 1826.
Presses Universitaires de France. Paris.

\qparr
Cristau (C.-A.) 1874: Du suicide dans l'arm\'ee. Thesis. Paris.

\qparr
Dickens (C.) 1842, 1996: Philadelphia and its solitary prison. Chapter 7 of
American Notes. Reedited by Modern Library. New York.

\qparr
Douglas (J.) 1967: The social meaning of suicide. 
Princeton University Press. Princeton.

\qparr
Drozdz (S.), Kwapien (J.), Gr\"ummer, Ruf (F.), Speth (J.) 2001:
Quantifying the dynamics of financial correlations.
Physica A 299,144-153.

\qparr
Halbwachs (M.) 1930: Les causes du suicide.
F\'elix Alcan, Paris.

\qparr
Hayes (L.M.), Rowan (J.R.) 1988: National study of jail suicides: seven
years later.
National Center on Institutions and Alternatives.

\qparr
H\o yer (G.), Lund (E.) 1993: Suicide among women related to number
of children in marriage.
Archive General of Psychiatry 50,134-136.

\qparr
HPMS (Her Majesty's Prison Service) annual report 2004, Appendix 1:
Statistical information. 

\qparr
Kim (K.), Yoon (S.-M.), Choi (J.S.), Takayasu (H.) 2004: Herd behaviors
in financial markets.
Preprint available on: http://arXiv.org/abs/cond-mat/0405172 (9 May).

\qparr
Krose (H.A.) 1906: Der Selbsmord im 19. Jahrhundert.
Herdersche Verlagshandlung, Friburg.

\qparr
MacDonald (M.), Sexton (S.) 2002: Self-harm and suicide policy:
implementation in West Middlands prisons. University of Central
England Report. Birmingham.

\qparr
Mantegna (R.N.) 1999: Hierarchical structure in financial markets.
The European Physical Journal B 11,193-197.

\qparr
Marshall (T.), Simpson (S.), Stevens (A.) 2000: Health care in prisons.
University of Birmingham Report (February 2000).

\qparr
Meer (F.) 1976: Race and suicide in South Africa. Routledge. London.

\qparr
Menezes (M.A. de), Barab\'asi (A.-L.) 2004: Separating internal and
external dynamics of complex systems. 
Preprint available on: http://arXiv.org/abs/cond-mat/0406421 (18 June).

\qparr
Nagle (J.T.) 1882: Suicides in New York cities during the 11 years ending
December 31, 1880. 
Riverside Press. Cambridge (Mass.).

\qparr
Nizard (A.) 1998: Suicide et mal-\^etre social.
Population et Soci\'et\'es 334, April.

\qparr
Pfeiffer (M.B.) 2001: Suicides high in prison box.
Poughkeepsie Journal [New York State], December 16, 2001.

\qparr
Pfeiffer (M.B.) 2002: Suicides in solitary confinement are abnormaly high.
Poughkeepsie Journal [New York State], April 14, 2002.

\qparr
Plerou (V.), Gopikrishnan (P.), Rosenow (B.), Amaral (L.A.N.),
Stanley (H.E.) 2001: Collective behavior of stock price movements:
A random matrix theory approach. 
Physica 299,175-180.

\qparr
Prison statistics: England and Wales 2001: Home Office. National
Statistics. London.

\qparr
Roehner (B.M.) 1995: Theory of markets. Springer. Berlin.

\qparr
Roehner (B.M.) 2005: A bridge between liquids and socio-economic
systems: the key role of interaction strengths. 
Physica A 348, 659-682.

Ropp (P.S.), Zamperini (P.), Zurndorfer (H.T.) eds 2001: Passionate
women: female suicide in late imperial China. 
Brill, Boston.

\qparr
Sattar (G.) 2001: Rates and causes of death among prisoners and 
offenders under community supervision. Home Office Research Study 231
[available on the website of the Home Office].

\qparr
Schimdtke (A.) et al. [26 co-authors] 1999: Suicide rates
in the world: update.
Archives of Suicide Research 5,81-89.

\qparr
Sornette(D.), Deschatres (F.), T. Gilbert (T.), Y. Ageon (Y.) 2004:
Endogenous Versus Exogenous Shocks in Complex Networks: an Empirical
Test Using Book Sale Ranking.
Physical Review Letters 93,22, 228701

\qparr
Stauffer (D.), Sornette (D.) 1999: Self-organized percolation model
for stock market fluctuations. 
Physica A 271, 496-506.

\qparr
Steenkamp (M.), Harrison (J.) 2000: Suicide and hospitalized self-harm in 
Australia.
Australian Institute of Health and Welfare. [Available on the Internet].

\qparr
Telzrow (M.E.) 2002: Punishment and reform: the Wisconsin State
reformatory. 
Voyageur, Winter-Spring, p. 25. [Available on the Internet].

\end{document}